# Triadic Social Structure Facilitates Backing for Crowdfunding Projects

Yutaka Nakai
Shibaura Institute of Technology

Faculty Systems Engineering and Science

Saitama, Japan

nakai@shibaura-it.ac.jp

Hiroki Takikawa
Tohoku University

The International Research Institute of Disaster Science (IRIDeS)

Sendai, Japan

takikawa@m.tohoku.ac.jp

*Abstract*—Crowdfunding is a new funding method through which founders request small amounts of funding from a large number of people through an online platform. Crowdfunding facilitates a new type of social capital and exhibits a unique form of social dynamics, thus attracting the interest of sociologists and other social scientists. Previous studies have focused on social relationships in crowdfunding such as direct reciprocity and consider how they contribute to the success of funding. The social structure of crowdfunding, however, involves more complex social relationships and it may contribute to the success of a new venture or project in many ways. In this study, we focus on a specific type of triadic social structure, the buddy relation, which can be described as a relationship through which project founder *x*, who previously backed another founder *z*'s project, receives financial backing from the other backers of *z*'s project. We found that the buddy relation occurs significantly more often than randomly, concluding that this structure facilitates the gathering of financial backing and may contribute to the success of a crowdfunded project.

## I. Introduction

Crowdfunding is a funding method through which founders, to realize their goals, request funds from crowd comprising many and unspecified individuals through an online platform. Crowdfunding sites are used for a variety of projects, such as video games, free software, inventions, scientific research, environmental initiatives, social welfare, and political activities. Founders make their own proposals in public on crowdfunding sites and ask backers to contribute a small amount of money. The founder sets up a target amount of funding and offers returns for their support such as products, thank you letters, or advertisements of backers' names as an acknowledgment. Individuals examine and compare proposals on the site. Those motivated select their favorite projects and decide the amount they want to pledge, and transfer money via micropayment. Finally, such backing expands via word of mouth on social networking services.

Backers have the following two types of motivation: extrinsic and intrinsic motivations. The former is based on returns, that is, self-interest, whereas the latter is based on sympathy toward founders [1]. For example, those motivated intrinsically see the feelings of "connectedness" to a community as precious, via participating social interactions [2].

The more attractive the proposal or return, the more fundraising is likely to be achieved. However, the major characteristic of crowdfunding is that new social capital emerges directly among people unknown to each other, and such social dynamics have attracted the interest of sociologists and other social scientists. Particularly, a large proportion of users in crowdfunding may interact with others not once but repeatedly; thus, their interactions form a type of social structure. Furthermore, present backers may become future founders and vice versa. This interdependent relationship creates the social dynamics of crowdfunding. For example, *b*'s previous experience of being backed by *a* may make *b* willing to back *a* this time; this is known as the reciprocity principle [3]. Social dynamics in crowdfunding, however, goes beyond simple reciprocity in crowdfunding, as discussed later in this study.

Previous studies on crowdfunding have mainly focused on factors contributing to funding success. They have examined the contents of a project and the way it is presented to identify elements that are crucial for success, such as the "goal" (the target amount of funding), "deadline," "news updates" (frequency of news updates), "Facebook likes" (number of Facebook likes), and length of explanation. Meanwhile, typographical errors usually result in failure [4].

Among the characteristics of crowdfunding, particularly interesting are those related to its social capital. References [4], [5], [6], and [7] explored the effects of founders' social capital on the success of a funding initiative. Reference [4] showed that the possibility of success is significantly correlated to the number of Facebook likes, which they regarded as a founder's social capital. Reference [5] showed that success is correlated to the social capital inside the crowdfunding site. As for the inside capital, [6] showed that a founder who has received backing previously from many individuals is likely to be able to collect more backing for future initiatives. Reference [7] also showed that a founder who has backed many others is more likely to receive more backing. Particularly, [7] have confirmed that direct reciprocity between founders exists in crowdfunding sites. As a related work, [8] showed that there is no indirect reciprocity among founders.

These studies shed light on the important mechanisms of social capital that make it possible to fundraise through crowdfunding sites. However, they do not show a more complex social structure than a dyad such as direct reciprocity (however, see [8] and the discussion). In this study, we explore such a social structure and its dynamics.

## II. THEORY

The questions explored in this study include the following: Why do backers back certain founders and not others? What is the effect of backing? How does previous backing by present founders contribute to gathering funds and thus the success of their projects?

The act of backing can be considered as the emergence of a direct tie between a backer and a founder. Thus, the backing-backed relationship forms a type of social network in crowdfunding sites. The network dynamics can be analyzed through social network theory [9] [10] [11] [12].

One of the most common theories is the theory of reciprocity; that is, if your project was backed by someone previously, then you should back her if she is seeking funding for her project. In fact, [7] showed the existence of direct reciprocity. However, direct reciprocity cannot be seen as the major dynamics because the number of founders is very less when compared to backers. Other mechanisms should be considered.

In social network theory, there are other candidates of mechanisms of tie creation apart from reciprocity. An important theory is that of triadic closure [13]. This theory states that you are likely to become friends with those who are friends of your friends. This dynamic has been found in various sites such as e-mail and social media [14] [15].

Furthermore, a shared focus tends to promote friendship between users. A focus here is defined as a "social, psychological, legal, or physical entity around which joint activities are organized (e.g., workplaces, voluntary organizations, hangouts, and families)" [16]. Reference [14] showed that class participation as a shared focus promoted friendship between college students. This is called focal closure.

Based on the ideas of triadic and focal closure, this study proposes a new hypothesis, different from direct reciprocity, on how founders' previous backing behaviors contribute to gathering funds. Our hypothesis is as follows: If $x$ backs project $P_z$, then backers of project $P_z$ (hereafter referred to as $w$) will be more likely to back $x$'s project $P_x$, as shown in Fig. 1.

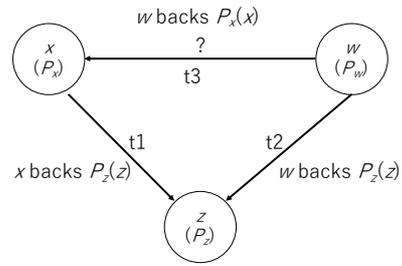

Fig. 1. Backers of project $P_z$ (hereafter referred to as $w$) will be more likely to back $x$'s project $P_x$, if $x$ backs project $P_z$.

This may be interpreted as the interaction of focal and triadic closures, as displayed in Fig. 2. First, this relationship may manifest focal closure because $P_z$ is a project in which the joint activity (backing the project) is organized. Specifically, when $x$ backs $P_z$, this backing relationship may express $x$'s shared concerns in the cause of the project launched by $z$. In the same way, $w$'s backing $P_z$ may indicate $w$'s shared concerns with $z$. Hence, it is expected that $x$ and $w$ may share similar concerns in the cause of the project that $x$ is trying to launch. In this way, $P_z$, as a focus, promotes shared concerns between $x$ and $w$.

However, only shared concerns may not guarantee a financial contribution to an unknown person. The $w$'s trust in $x$ will be also required. Here, it must be noted that founder $z$ may not only be a shared focus for $x$ and $w$ but may also be a mutual friend, which implies that a type of triadic closure may also be involved in this relationship. As a mutual friend, $z$ may promote this trust relationship. We assume that this is attributed to the following story. When $z$ knows that $x$ is launching her project $P_x$, $z$ informs $w$ of this and asks $w$ to back $x$'s project $P_x$. Subsequently, $w$ may be convinced that $x$ and her project $P_x$ is trustworthy because her friend $z$ has guaranteed it.

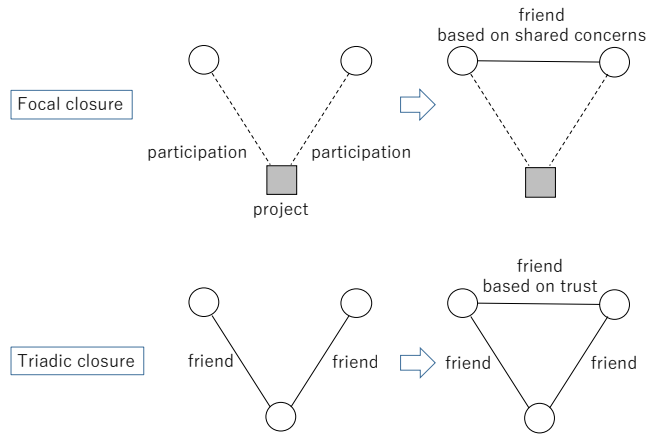

Fig. 2. Focal closure and triadic closure

When this interaction of two mechanisms activates a friendship between $x$ and $w$, and $w$ backs $x$'s newly launched project $P_x$, we refer to this as the "buddy effect."

This is not merely our speculation. We interviewed the founder of a crowdfunding site[1] to confirm whether the above relationship is observed often. He told us that successful projects typically gather 1/3 of their funding from the founder's direct friends, 1/3 from friends of friends, and 1/3 from others. Furthermore, he stated,

> *As a rule of thumb, what a founder should do is not to ask for contributions from her direct friends, but to say to them, "Please let your friends know that I am launching my project."*

---

[1] We interviewed D.S., who is the founder of a popular crowdfunding site in Japan. We conducted the interview on June 27, 2018.

This clearly refers to the process of closure that we hypothesized. However, this is just evidence; thus, quantitative evidence must be provided to verify this hypothesis, which we will do later in the study. His story suggests that our theoretical hypothesis is not merely speculative but worth exploring.

In sum, our theory predicts that if $x$ backed $P_z$, which was backed by $w$, then $x$ is more likely to be backed by $w$ in the future. As a corollary, if $x$ backed $P_z$ and a founder $z$ has many backers, then $x$ can gather more funding.

## III. DATA

### A. Data Collection and Organization

Crowdfunding consists of four categories [17]: donations [18], rewards [4][5][19], debts [20], and equity [21][22]. Donations do not give backers any returns, whereas rewards, debts, and equity give backers non-monetary returns (e.g., a product or a thank you letter), dividends, and interest, respectively. Donations and rewards represent an essential aspect of crowdfunding because they have nothing to do with money. In the study, we collected data for these crowdfunding categories, particularly from Readyfor, Japan's largest crowdfunding site. The data were collected from the site's activities spanning from May 16, 2011 to September 5, 2017. The data are organized into the following two classes: user data and project data. The user data were collected from 306,968 users. Several variables were included such as user ID, username, and user description. While users can be founders, they are not necessarily founders in this case; the majority of Readyfor users only provide backing and do not launch their own projects. We will refer to founders as the "community," and users who only back others' projects as the "crowd." The project data consists of 6,559 different projects. The project information includes project ID, founder ID, project category, who commented on the project and when, and project deadline. Comment information is critical because it can be used to identify backing relationships and the dates of these links. In Readyfor, it is a norm to leave comments if you back a project. Since the commenting time is recorded, this can be used as a proxy for the time of the occurrence of the backing event. Project deadline is another valuable source of information. It specifies when the call for project funding ended. The problem is that it is impossible to specify the starting time of a project. As a proxy for this, we used the time of the first comment on a project as its starting time.

By combining user and project data, we obtained a bipartite network whose nodes are projects and users and whose edges are backing relations from users to projects, which is shown in Fig. 3. The number of backer nodes is 203,568 (this is smaller than the total number of users because the latter includes users who do not back at all). The number of project nodes is 6,559, and the number of edges is 279,676.

There are two points to note in this case. First, project nodes have their founders. This implies that the user-project relationship also represents the user-founder relationship. Second, this is longitudinal network data; that is, edges have time stamps, and project life spans (from cut off to deadline) are also recorded. This time information is critical for our analysis.

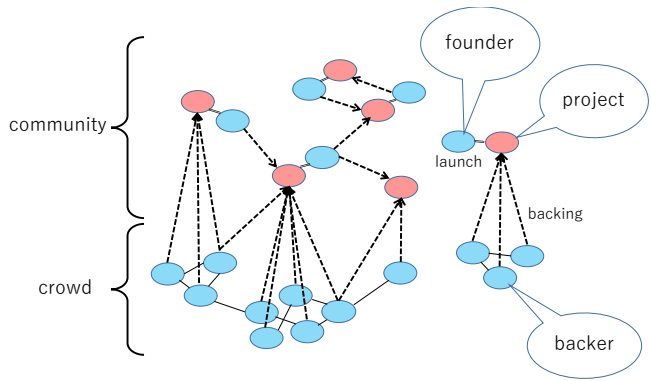

Fig. 3. The bipartite network of users and projects in Readyfor

### B. Basic Statistics and Preliminary Analysis

The bipartite network created from backing relations in Readyfor had a very skewed degree of distribution, as expected. The basic statistics of project in-degree distribution are shown in Table I and Fig. 4.

TABLE I. BASIC STATISTICS OF PROJECT IN-DEGREE DISTRIBUTION

| Mean | Std | Min | 25% | 50% | 75% | Max | Mode |
|------|-----|-----|-----|-----|-----|-----|------|
| 42.6 | 75.3 | 0 | 5 | 21 | 53 | 1402 | 0 |

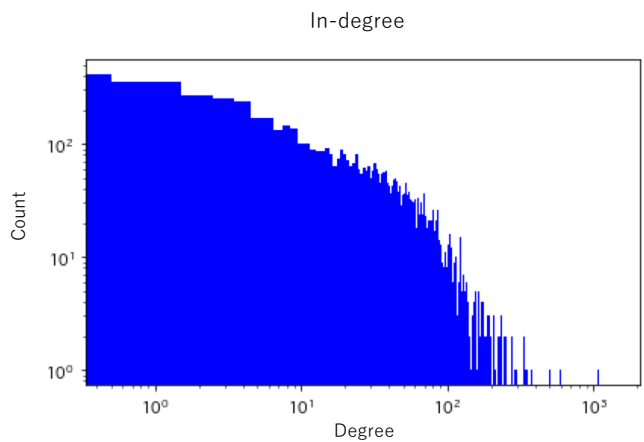

Fig. 4. Log-log plot for project in-degree distribution

The mode of this distribution is 0. Specifically, 411 projects received no backing throughout their life spans. On the other hand, less than 25% of the projects obtained support from over 50 different backers, with the highest number reaching 1,402. This shows the skewedness of project in-degree distribution.

The user out-degree distribution was much more skewed, as shown in Table II and Fig. 5.

TABLE II. BASIC STATISTICS OF USER OUT-DEGREE DISTRIBUTION

| Mean | Std | Min | 25% | 50% | 75% | Max | Mode |
|---|---|---|---|---|---|---|---|
| 1.4 | 3 | 1 | 1 | 1 | 1 | 733 | 1 |

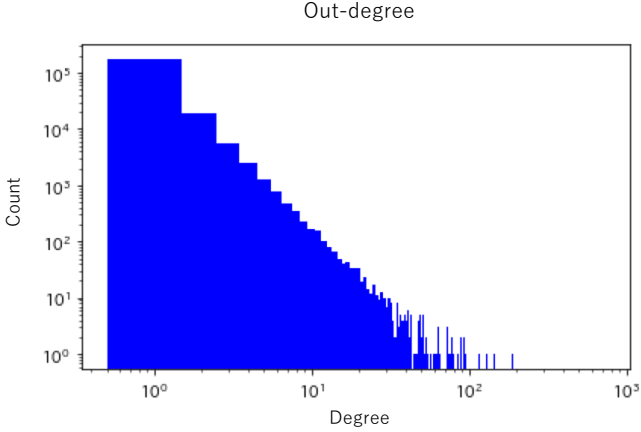

Fig. 5. Log-log plot for user out-degree distribution

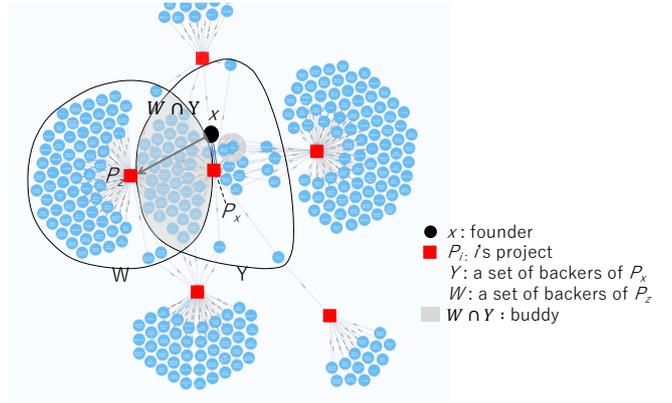

Fig. 6. An example of the buddy relation

In fact, about 85% backers (172,232 backers) backed someone's project only once. The number of two-time and three-time backers was 19,006 and 5,705, respectively. These numbers account for more than 96% of the backers. Very few people backed projects more than ten times (approximately, 0.4% of the total backers).

As the next step, we examined whether there was any prima facie evidence of the buddy effect. A buddy relation is observed in the following two conditions.

1. Both $x$ and $w$ backed $P_z$. The backer that backed first ($x$ or $w$) is irrelevant.
2. After condition 1 is satisfied, $w$ backs $P_x$.

We can calculate the buddy ratio by dividing the number of case 2s by the number of case 1s.

Buddy ratio = the num. of cases where $w$ backs $P_x$ / the num. of cases where both $x$ and $w$ backed $P_z$

If the buddy ratio in Readyfor is significantly high, then it can be considered as prima facie evidence for the existence of the buddy effect in this platform.

First, we calculated the denominator of the buddy ratio, that is, the number of cases where both $x$ and $w$ backed $P_z$. The mean was 106.5. It must be noted that we excluded the 0-case in calculating the mean because when the denominator is 0, the buddy ratio cannot be calculated. This means that if you ($x$) are not the only backer of someone's ($z$'s) project $P_z$, there are other 106.5 backers ($w$) of $P_z$, on an average, that are potential candidates for backing your project $P_x$ through the buddy relation. In real data, out of 106.5, 1.7 people backed $P_x$, on an average. Thus, the buddy ratio is 0.031.

Fig. 6 shows a typical example of a buddy relation. $Y$ and $W$ are a set of backers of $P_x$ and $P_z$, respectively. $P_x$ is backed by $W \cap Y$ who backed $P_z$, which was also backed by X.

Although the buddy relation appears to exist here, it is difficult to determine whether the raw ratio 0.031 indicates the existence of the buddy effect. The number may be generated through a purely random process. If $w$ has a large out-degree, then $w$ may back $P_x$ randomly without knowing that $x$ backed the same $P_z$ as itself. However, this relation is created purely accidentally and is not considered a genuine buddy relation. To exclude this possibility, a more sophisticated method must be employed.

## IV. METHOD

To confirm whether the buddy effect works, we conducted a conditional uniform graph (CUG) hypothesis test [10]. In a CUG test, the null hypothesis is that the observed graph is uniformly generated and conditional on the assumed properties of nodes and edges. Under this hypothesis, many simulated graphs are generated via the Monte Carlo simulation. Subsequently, the statistics between the observed graph and simulated graphs are compared to examine whether there are any significant differences.

For our null hypothesis, we must erase only the fact that $w$ backed $P_z$ while preserving backers' tendencies to back and projects' tendencies to attract backings. First, to preserve the backers' tendencies, backers' out-degrees were fixed; if backers $a$, $b$, and $c$ backed 4, 8, and 2 times, respectively, in the observed network, then their out-degrees in the simulated network should remain the same at 4, 8, and 2, respectively. Second, to preserve the projects' tendencies, backers choose projects to fund according to the "popularity" of the projects. The popularity in this sense is defined as the number of backers the project gathers in the real world. Thus, if project $P_a$ gathered more backing than project $P_b$ in the real world, project $P_a$ is more likely to be chosen as a backing destination than project $P_b$. More details concerning this condition are provided below.

The Monte Carlo simulation runs as follows. As we shall see below, all edges in the bipartite graph are randomly rewired according to the conditions mentioned above. Let us define the edge from backer $a$ to project $P_b$ as $e_{aPb}$, which has its own time stamp. Next, we must define a set of candidates of projects that are the target for rewiring as $CS$ and denote $|CS|$ as $K$. The actual $e_{aPb}$ is rewired and $e_{aPk}$ (where $P_k \in CS$) is created in the simulated network. $P_k$ is chosen probabilistically from $CS$—the set of candidates. $CS$ is subject to the time constraint—all projects that exist when backer $a$ backed project $P_b$ are assigned as candidates. In other words, the life span of the candidate should include the creation time of $e_{aP_b}$. The assumption here is that backers would choose among the candidate projects that run simultaneously with project $P_b$ at the time of backer $a$ backing it in the real world. Thus, the number of elements of *CS* as rewiring targets is much smaller than the number of total project nodes in the graph.

How is $P_k$ chosen from *CS*? This process should be formalized by the popularity of projects. The in-degree of $P_k$ is denoted as $s_k$. $s_k$ is the number of backers that the project $P_k$ gathered in the real world. In the simulation, the backer chooses the backing destination probabilistically, according to the categorical distribution,

$$Cat(d|\mu) = \prod_{k=1}^{K} \mu_k^{d_k},$$

where $d_k$ is the element of k-dimensional vector $d$ and its value is 0 or 1. Furthermore, $\sum_{k=1}^{K} d_k = 1$. In other words, $d_k$ is the indicator of $P_k$ being chosen as the backing destination. $\mu_k$ is the probability of being chosen, which is defined as $s_k / \sum_{k=1}^{K} s_k$. In other words, $P_k$ is chosen with the probability of $P_k$'s in-degree divided by the sum of in-degrees of all the candidates for the rewiring targets.

Let us explain the procedure using a hypothetical example illustrated in Fig. 7. Consider rewiring the edge from $a$ to $P_b$. This backing occurred on July 1, 2016. Assume that a set of candidates for the rewiring targets consist of $P_i$, $P_j$, $P_k$, and $P_l$. It must be noted that life spans for every project include the date when the backing from $a$ to $P_b$ occurred. The rewiring probability is determined in terms of each in-degree, that is, the rewiring probability for $P_i$ is 4/25, $P_j$ is 7/25, $P_k$ is 12/25, and $P_l$ is 2/25. In this case, the rewiring happened from $a$ to $P_k$.

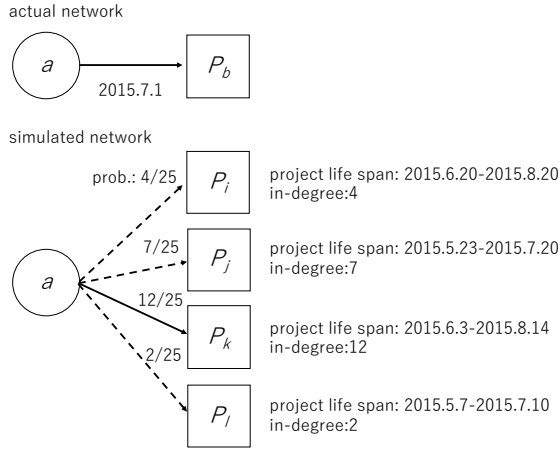

Fig. 7. A hypothetical example of the simulated network

In sum, this simulation satisfies the two conditions discussed earlier. First, since each actual edge is rewired one by one, the out-degree distribution remains unchanged. Second, rewiring probability follows the categorical distribution, reflecting the popularity of each project. We conducted this simulation 100 times.

V. RESULTS

For each simulated network, we calculated the buddy ratio defined above. The distribution of the ratio over 100 trials is shown in Fig. 8. The mean of the ratio is 0.0011.

Since the actual buddy ratio is 0.031, the null hypothesis that this occurs under the above two conditions is correct, with a probability of less than 0.01. Thus, it was concluded that the buddy effect exists in the crowdfunding site Readyfor.

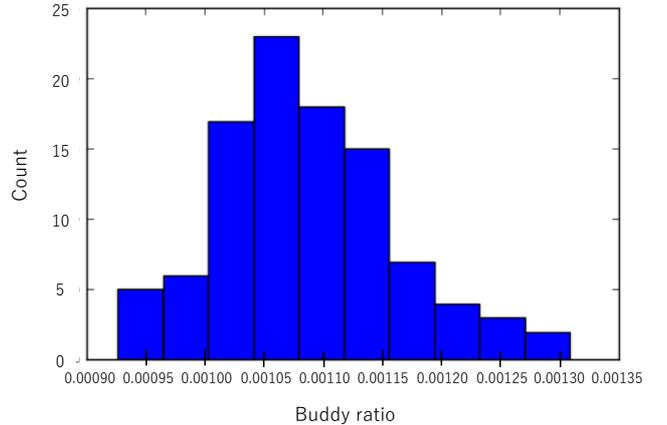

Fig. 8. The distribution of the buddy ratio over 100 trials

VI. DISCUSSION

This study proposed the buddy relation as a more complex structure than a dyad to explain the facilitation of backing for projects. As a similar work, [8], focusing on a transitive triplet, which is a type of hierarchical network (Fig. 9), showed that the triplets significantly exist in the network among founders of Kickstarter, the leading crowdfunding site in the world. Although the transitive triplet and buddy relation appear to have similar structures (Fig. 9), the two are different in the following ways. First, all nodes of the transitive triplet, $x$, $z$, and $w$, are founders, while the $w$ from the buddy relation are pure backers. Second, the edges of the triplet, $xz$, $wz$, and $wx$, have no ordering in time because network data was aggregated over time, while the order of edges of the buddy relation is that $wx$ occurs after $xz$ and $wz$. Finally, this study showed fine micro structures among founders and backers, which reflects a causal relationship.

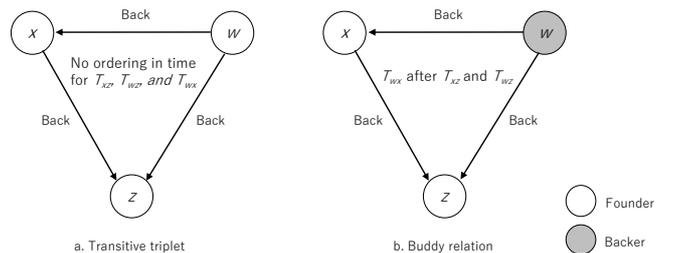

Fig. 9. Transitive triplet and buddy relation.

There are some limitations to the present study. First, the study focused on a single platform; thus, the generalizability of the finding is open for debate. A cross comparison of different crowdfunding platforms is a possible field for future research. Second, the study found that the buddy relation worked effectively, but this did not exclude the possibility that other social mechanisms are also effective in gathering the funding. For our next study, we aim to consider other mechanisms and compare their strengths with the buddy effect. Finally, although we showed that the buddy relation increased the probability of receiving backing, and this indicates that it may contribute to the success in gaining funds eventually, direct evidence for this possibility was not provided. This factor should be demonstrated through future research.

## VII. CONCLUSION

Crowdfunding can be a significant force of social change in modern society. Generally, cultural, environmental, and welfare issues have difficulty gaining funding because their marketability is poor (the failure of the market), and they do not serve as vote-gathering mechanisms (the failure of government). Conversely, besides marketing (the self-aid) and governmental aid (the public aid), crowdfunding serves to provide funding to social entrepreneurs, as a form of mutual aid. As a result, crowdfunding can be regarded as a new institution responsible for the redistribution of resources; thus, crowdfunding has a high social significance. However, the transfer of money between people who have no existing or weak relations with each other is quite difficult. To establish trustworthy relations among people beyond direct reciprocity, both the triadic and focal closure conditions need to be present. The performance of a joint activity to support a project and the existence of mutual friends provide opportunities for the creation of sympathetic feelings and trust building. We found evidence that such an essential dynamic, the buddy effect, exists and helps founders to gain funding for their projects. However, the ratio of 0.031 is quite small. The next generation of crowdfunding should facilitate the establishment of the buddy relation.

ACKNOWLEDGMENTS

This work was supported by JSPS KAKENHI Grant Numbers JP16H03698、JP16K04027, and 18H03621. The authors give special thanks for the fruitful discussions to Prof. Hiroshi YAMAMOTO (Rissho University), Mr. Masaya OOTANI, Mr. Takehiro ARAI, Mr. Toshiki AMEMIYA, and Prof. Yosuke KIRA (University of Aizu), who passed away on 16 September, 2018.